\documentclass[11pt]{article}

\usepackage[margin=1in]{geometry}
\usepackage{amsmath,amssymb,amsfonts}
\usepackage{booktabs}
\usepackage{graphicx}
\usepackage{algorithm}
\usepackage{algpseudocode}
\usepackage{tikz}
\usetikzlibrary{positioning,arrows.meta,calc}
\usepackage{listings}
\usepackage{xcolor}
\usepackage{url}
\usepackage[hidelinks]{hyperref}

\newcommand{\tile}{\texttt{tile\_n}}
\newcommand{\code}[1]{\texttt{#1}}

\title{\bf GC-MoE on SM120:\\ Overlapping Expert-Parallel Token Dispatch with\\ Warp-Level FP8$\times$FP4 Block-Scaled MMA on Consumer Blackwell}

\author{%
\begin{tabular}{c}
Yong Li \quad Haoyu Wang \quad Kan Liu \quad Tao Lan\\[3pt]
Alibaba Inc.\\[2pt]
\texttt{\{brucelee.ly, why492961, liukan.lk, tao.lant\}@alibaba-inc.com}
\end{tabular}}
\date{}

\begin{document}
\maketitle

\begin{abstract}
Expert-parallel (EP) Mixture-of-Experts (MoE) decoding is a latency problem before it is a
throughput problem: every layer serializes a token exchange over the interconnect, a
grouped GEMM over the routed experts, and a combine, while only a few tokens per rank are
in flight. Prior GC-MoE work attacks this serialization on Hopper (SM90) and datacenter
Blackwell (SM100) by co-scheduling the communication and the compute of one layer inside
one persistent, Green-Context-partitioned CUDA Graph.

This paper describes GC-MoE for \emph{SM120} (compute capability 12.0, e.g.\ RTX PRO 5000),
the consumer/workstation Blackwell variant. SM120 is not a mechanical port of the
datacenter design. It exposes block-scaled tensor cores only through warp-level
\code{mma.sync.aligned.kind::mxf8f6f4}, so there is no tensor memory, no two-CTA MMA and no
cluster-wide multicast to lean on; FP4 weight nibbles must be presented to the MMA per
warp; the SM budget that can be partitioned between communication and compute is small; and
the inter-GPU fabric is PCIe. We therefore rebuild the kernel around a \emph{swap-AB}
warp-level block-scaled MMA in which FP4 weights occupy the register $A$ operand and FP8
activations the $B$ operand, and we add a set of SM120-specific mechanisms: packed
block-scale byte selection through MMA immediates that removes all scale-byte shuffling from
the inner loop; a compact-weights path that halves weight staging and expands FP4 in
registers; a run-time adaptive token-tile geometry that selects among register/SMEM/global
handoffs; an expert-granular streaming dispatch with a batched bulk-pull engine, a
64-bit pull cache that deduplicates activation reads by (source rank, token), and
packed-route and mirrored-metadata exchange; full-width BF16 returns that are coalesced into
bulk stores on the direct path and drained by resident relay copiers over a position-major
arena; and shared experts folded into the compute role, which reuses the routed role's
shared-memory budget instead of occupying a kernel of its own. On 8$\times$RTX PRO 5000 over
PCIe, the resulting operator is on average $1.522\times$ faster than the FlashInfer CuTe DSL
W4A8 MegaMoE and $1.387\times$ faster than the Cake MegaMoE across 18 decode points and two
model shapes (18/18 wins each), and $1.388\times$ faster than Cake on the fused
shared-expert variant; all 94 numerical cases and 752 per-rank checks pass, with
cross-operator cosine similarity of at least $0.9999997$.
\end{abstract}

\section{Introduction}

A Mixture-of-Experts (MoE) layer replaces one dense feed-forward network with $E$ experts and
a router that sends each token to its top-$k$ experts. Under expert parallelism (EP) the
experts are partitioned across ranks, so a single layer performs three dependent phases:
(i) \emph{dispatch}, an all-to-all-like exchange that moves each token's activation to the
ranks owning its selected experts; (ii) a \emph{grouped GEMM} over the experts, computing
the gate/up projection with a fused SwiGLU and then the down projection; and (iii) a
\emph{combine}, which returns each expert's contribution to the token's owner and weights and
sums the $k$ contributions.

At decode time the batch per rank is tiny---one to a few hundred tokens---so the layer is
latency bound and badly under-utilized in each phase separately. The motivating observation
of GC-MoE~\cite{gcmoe} is that the three phases do not need to be separated by global
barriers: once a peer's token row has landed, the experts that consume it can start, and the
combine of one expert tile can proceed while later experts are still multiplying. GC-MoE
therefore fuses dispatch, expert computation and combine into one \emph{three-node CUDA
Graph} whose nodes are pinned to disjoint SM partitions created with CUDA Green
Contexts~\cite{greenctx}: a communication partition runs the dispatch, a compute partition
runs the grouped GEMM, and a small finalize partition performs the combine, all replaying
per layer.

That design was built for Hopper and for datacenter Blackwell, and it leans on hardware those
parts have: TMA multicast and distributed shared memory within a cluster, the fifth-generation
tensor core with tensor memory (\code{tcgen05}) and two-CTA collaborative MMA, and
NVLink-class interconnect bandwidth. The subject of this paper is the same operator on
\emph{SM120}, the compute capability that powers the RTX PRO 5000 and similar
workstation/consumer Blackwell parts, connected over PCIe. Porting the SM100 kernel there
fails for four structural reasons.

\paragraph{(1) The tensor-core programming model changes.} SM120 does expose block-scaled
(``microscaling'') tensor cores, but only through the warp-level
\begin{center}
\code{mma.sync.aligned.kind::mxf8f6f4.block\_scale.scale\_vec::1X.m16n8k32.row.col}
\end{center}
instruction, which we call \emph{QMMA}. There is no \code{tcgen05.mma}, hence no tensor
memory, no asynchronous MMA with a separate completion barrier, no two-CTA collaborative
MMA, and no cluster-wide activation multicast. A $64\times N\times K128$ tile must be
assembled out of $m16n8k32$ atoms issued by four math warps, and every operand must reach
registers through SMEM.

\paragraph{(2) FP4 weights must be manufactured per warp.} Routed weights are MXFP4
(E2M1 with UE8M0 scales), i.e.\ half a byte per element, while activations are FP8 (E4M3).
The mixed-input QMMA consumes both operands from registers, so the 4-bit payload has to be
turned into a legal operand: either expanded to 8-bit during the global-to-shared transfer, or
staged as raw nibbles and expanded in registers. Both options exist in our kernel, and the
choice interacts with SMEM capacity, which in turn decides how many CTAs can co-reside on the
compute partition.

\paragraph{(3) The partition is small, so geometry must adapt.} With six SMs defaulted to
dispatch, the compute partition must still hold at least 32 SMs and an even count, and the
whole compute grid (including the CTAs that stream results back) must be \emph{resident}---the
role kernels synchronize grid-wide. There is no room for a fixed, generously provisioned tile.

\paragraph{(4) The interconnect is PCIe.} A routed activation row costs $H$ bytes of FP8 plus
$H/32$ bytes of packed scales; its contribution returns as BF16, i.e.\ $2H$ bytes. Over PCIe
the return leg is therefore the single largest term of the exchange, and it is the term that
the datacenter designs treat as free.

This paper presents the SM120 GC-MoE operator (GC MoE 0.6.0, commit \code{141d287}), which
addresses these four points with new mechanisms rather than retuning. Our contributions are:

\begin{itemize}
\item \textbf{Swap-AB block-scaled mainloop.} We place the FP4 weights in the QMMA $A$
operand and the FP8 activations in $B$. This keeps the weight tile ($M{=}64$ rows,
$K{=}128$) invariant and lets the token extent---the $N$ dimension---change at run time
without touching the weight pipeline, which is what makes the adaptive geometry of
\S\ref{sec:geom} cheap. We also feed the QMMA \emph{packed} scale words and select the
per-$K32$ scale byte with the instruction's byte-id immediates, eliminating the
shift/permute scale bookkeeping from the persistent loop (\S\ref{sec:qmma}).
\item \textbf{Two FP4 operand paths.} A 4D TensorMap expands FP4 to 8-bit during the TMA; a
compact path instead keeps the raw nibble tile at half the bytes and expands it with four
\code{ld.shared.u16} plus \code{prmt} per fragment. The geometry selects between them
(\S\ref{sec:qmma}, \S\ref{sec:geom}).
\item \textbf{Adaptive geometry with global and shared handoffs.} Token-tile width, pipeline
depth, weight compactness, L1 handoff placement (SMEM vs.\ global vs.\ shared 32-row chunks),
co-residency and register ceiling are derived from the routed-row count and the SMEM budget at
launch time (\S\ref{sec:geom}).
\item \textbf{Streaming, expert-granular dispatch.} Experts are published the instant their
last row lands; peer rows are pulled with a batched \code{cp.async.bulk} engine that keeps
several PCIe reads in flight; a producer-side pull cache keyed by (source rank, token) turns
the several local experts a single token usually reaches into one PCIe read, and publishes
the local row address through a single 64-bit acquire/release word; routes travel as a packed
array and activation scales and route weights are mirrored into the peer workspace so the
compute partition reads them locally (\S\ref{sec:dispatch}).
\item \textbf{Full-width BF16 returns with resident relay copiers.} The return leg is twice
the forward payload and is the largest term of the exchange on PCIe, so the operator spends
its effort on how those bytes move rather than on a custom codec: the direct path coalesces
consecutive destination rows of a tile into one bulk store with a single copy group in
flight, while the relay path parks every finished row in a position-major arena that a
resident copier CTA drains with double-buffered 16\,KB transfers, following the FC2 active
queue so a slow expert cannot hold up output that is already complete (\S\ref{sec:return}).
\item \textbf{Shared experts inside the compute role.} Rather than launching FC1/FC2 as
separate kernels, the compute role runs them on its first four warps against the same
dynamic shared memory the routed mainloop uses, ordered by per-token-tile completion
counters and a named barrier, so the shared path costs neither an extra role nor extra
residency (\S\ref{sec:se}).
\end{itemize}

On 8$\times$RTX PRO 5000 (PCIe) the operator wins all 18 decode points of two model shapes
against both baselines, averaging $1.522\times$ over CuTe DSL and $1.387\times$ over Cake for
the routed operator and $1.388\times$ over Cake with the fused shared expert, and passes
94/94 numerical cases.

\section{Background}

\subsection{MoE decoding under expert parallelism}

Let a rank own $L = E/R$ local experts, let $B$ be the number of tokens resident on a rank
per decode step, and let $k$ be the routing fan-out. Each rank contributes $Bk$ assignments to
the layer and computes $Bk$ of them on average, so the layer moves $RBk$ assignments in total.
The assignments a rank receives are spread over its $L$ local experts, which fixes the routed
row count per expert,
\begin{equation}
r \;=\; \Big\lceil \tfrac{B\,k}{L}\Big\rceil ,
\label{eq:rows}
\end{equation}
is the quantity that decides the GEMM tile shape. Decode batches make $r$ small: with
$L{=}32$, $k{=}8$ and $B{=}1$ (\S\ref{sec:eval}), $r{=}1$; even at $B{=}256$ we have
$r{=}64$, which the geometry of \S\ref{sec:geom} covers with a single $\tile{=}80$ tile. These
are token-side rows: in the swap-AB orientation of \S\ref{sec:qmma} they are the $N$ extent,
while the $M{=}64$ extent belongs to the weights.

For every remote assignment the exchange moves the activation row ($H$ bytes of FP8), its
microscaling factors ($H/128$ int32 words, i.e.\ $H/32$ bytes) and, after the down
projection, a BF16 partial sum ($2H$ bytes). Thus the per-assignment wire volume is
\begin{equation}
v \;=\; H + \tfrac{H}{32} + 2H,
\label{eq:volume}
\end{equation}
which on PCIe makes the return leg twice the forward leg. The SM120 operator keeps this
payload intact and attacks the term from the other side: it removes redundant reads of the
forward leg (\S\ref{sec:dispatch}), and it makes the return leg a bulk-copy problem handled
by CTAs that are not doing arithmetic (\S\ref{sec:return}).

\subsection{GPU execution model}

\paragraph{Green Contexts.} CUDA 12.4+ exposes driver-level SM partitioning: a device resource
is queried with \code{cuDeviceGetDevResource}, split by SM count with
\code{cuDevSmResourceSplitByCount}, and turned into an independent execution context with
\code{cuGreenCtxCreate}. A \emph{balanced workqueue} resource
(\code{CU\_WORKQUEUE\_SCOPE\_GREEN\_CTX\_BALANCED} with concurrency limit 1) guarantees that
concurrently resident green contexts make forward progress instead of one starving the other.
GC-MoE uses exactly this: the dispatched partition and the compute partition are separate
contexts with separate streams, so a spinning consumer cannot block the producer.

\paragraph{Persistent role graphs.} Each layer is a CUDA Graph of kernel nodes whose
dependencies are declared explicitly. Because the role kernels synchronize grid-wide through
epoch-tagged counters in device memory, every CTA of a role must be co-resident with the rest
of its role; the runtime therefore validates each role's grid against the occupancy of its
partition before instantiating the graph.

\paragraph{Scale formats.} An MXFP4 weight is a 4-bit E2M1 code with a shared UE8M0 exponent
per 32 contiguous elements; an MXFP8 activation uses E4M3 with the same block structure. The
QMMA multiplies a 4-bit operand by an 8-bit operand, adds into FP32, and applies both
block scales. A $K{=}128$ reduction therefore needs four $K32$ blocks, i.e.\ four scale bytes
per lane.

\section{Motivation: what SM120 changes}

With the dispatch overlapped against compute, the steady-state cost of a layer is
\begin{equation}
T_{\text{step}} \;=\; t_{\text{scan}} + t_{\text{pro}}
   + \max\!\big(T_{\text{pull}},\, T_{\text{gemm}}\big)
   + \max\!\big(T_{\text{ret}},\, t_{\text{clean}}\big)
   + t_{\text{tag}},
\label{eq:cost}
\end{equation}
where $t_{\text{scan}}$ is route scanning, $t_{\text{pro}}$ the time to publish the first
expert, $T_{\text{pull}}$ the PCIe pull time of the activation stream, $T_{\text{gemm}}$ the
grouped GEMM time, $T_{\text{ret}}$ the return traffic, $t_{\text{clean}}$ the SMEM/global
recycling, and $t_{\text{tag}}$ the completion handshake. Compared with the SM90/SM100 case,
each term changes character.

\paragraph{MMA efficiency.} One QMMA retires $16\times 8\times 32$ multiply-accumulates from
registers. To keep the tensor pipe busy the inner loop must issue back-to-back, which leaves
no budget for scale bookkeeping. On SM100 the scale factors are gathered by the tensor core
from SMEM; on SM120 they are ordinary register operands, so any per-issue byte extraction
(\code{shf}, \code{lop3}, \code{prmt}) directly steals MMA issue slots. Our first design
decision follows: keep the four $K32$ scale bytes of a $K128$ block in \emph{one} 32-bit
register and let the instruction choose the byte (\S\ref{sec:qmma}).

\paragraph{Operand orientation.} If the token dimension is placed on $M$, the activation
tile's SMEM footprint and its legal TMA descriptor change with the token count, and the
weight operand would have to be transposed (which the FP4 matrix load cannot do). Placing the
weights on $M$ and the tokens on $N$ decouples the two: $M{=}64$ is fixed, and the only
run-time quantity is how many $N{=}8$ groups are issued.

\paragraph{SM budget and co-residency.} The compute partition holds the MMA CTAs \emph{and}
the CTAs that stream returns, all resident. The number of co-resident CTAs per SM is
\begin{equation}
c \;=\; \max\!\Big(1, \min\big(4,\, \big\lfloor S_{\text{sm}} / S_{\text{role2}}\big\rfloor\big)\Big),
\qquad
R_{\text{cap}} \;=\; 16\,\Big\lfloor \frac{65536}{192\,c\cdot 16}\Big\rfloor ,
\label{eq:ctas}
\end{equation}
where $R_{\text{cap}}$ is the per-thread register budget rounded down to the allocator's
16-register granularity (and left unset when only one CTA co-resides). A few kilobytes of SMEM
therefore decide whether the kernel runs one CTA per SM (and loses half its latency hiding) or
four. Geometry must therefore be selected per shape, not per architecture.

\paragraph{PCIe and the return leg.} By \eqref{eq:volume} the return leg is $2\times$ the
forward leg per assignment. At $B{=}128$ with $k{=}8$ over 8 ranks (GLM5.2) the measured
payload rate is only $24$\,GB/s per GPU (\S\ref{sec:eval}), far below the memory system, so
the return path is where the operator must be frugal in \emph{transactions} rather than in
bytes: one bulk copy per contiguous run, and copiers that do not steal MMA CTAs.

\paragraph{Small batches.} At $B\!\le\!2$ the whole schedule is dominated by control: the
scan, the publish handshake and the completion tag. There, an alternative dispatch path that
scans peers directly from one CTA and opens a local gate for the rest beats the general
pipeline.

\section{The SM120 GC-MoE Design}

\subsection{Overview}

The layer is executed by three role kernels. A role is a (grid, threads, SMEM) tuple launched
into one of two Green contexts; role 1 runs in the dispatch context and roles 2 and 3 in the
compute context. Table~\ref{tab:roles} lists them and
Algorithm~\ref{alg:layer} states how the graph is composed. There is no fourth role: the
shared expert of \S\ref{sec:se} is compiled into role 2, and the return leg of
\S\ref{sec:return} is served by role-2 CTAs and by the utility warp of the mainloop, so the
fused variant runs the same three kernels with a different argument vector.

\begin{table}[t]
\centering
\small
\begin{tabular}{@{}llll@{}}
\toprule
Role & Threads & Context & Work \\
\midrule
1 & 256 & dispatch (6 SMs) & scan routes, pull peer rows, publish experts \\
2 & 192 & compute & grouped GEMM1+SwiGLU, GEMM2, return staging, relay, \\
  &     &         & shared-expert FC1/FC2 \\
3 & 384 / 256 & compute & combine $k$ partial sums, BF16 store of $y$ \\
\bottomrule
\end{tabular}
\caption{SM120 role kernels. Role 3 runs with 384 threads on the unbatched path and with 256
once returns are batched through the relay arena. The copiers are role-2 CTAs: the blocks
below $G_{\text{relay}}$ stream the arena instead of running the MMA.}
\label{tab:roles}
\end{table}

\begin{algorithm}[t]
\caption{Layer replay (SM120)}
\label{alg:layer}
\begin{algorithmic}[1]
\State $\triangleright$ persistent, replay-aware counters; no per-layer reset
\State epoch $\leftarrow$ epoch $+1$ in the header word (release)
\State launch role 1 with a programmatic-dependency edge to role 2
\For{each role-2 CTA $b$}
  \If{$b < G_{\text{relay}}$}
    \State drain the position-major relay arena to the peer it owns
  \Else
    \If{shared expert is fused}
      \State warps $< 4$: FC1/FC2 tiles, ordered by completion counters
    \EndIf
    \State per expert: claim $\rightarrow$ W/X pipeline $\rightarrow$ L1 $\rightarrow$ L2
  \EndIf
\EndFor
\State role 3 depends on all compute nodes; combines and writes $y$
\end{algorithmic}
\end{algorithm}

\begin{figure}[t]
\centering
\begin{tikzpicture}[font=\footnotesize,
  ctx/.style={draw=black!45, rounded corners=3pt, fill=black!3, align=center, inner sep=3pt},
  bar/.style={draw=black!40, fill=black!14, minimum height=0.26cm, inner sep=1pt,
              font=\scriptsize, anchor=west},
  ar/.style={-{Latex[length=1.7mm]}, semithick, black!60},
  head/.style={font=\scriptsize, text=black!60}]

\node[ctx, text width=2.95cm, minimum height=2.05cm] (r1) at (0,0)
  {\textbf{role 1}\\ dispatch\\[3pt]
   {\scriptsize 256 threads}\\[3pt]
   {\scriptsize scan routes\\ bulk pull\\ (8 rows in flight)\\[2pt] publish per expert}};
\node[head, above=2pt of r1] {dispatch context, 6 SMs};

\node[ctx, text width=5.30cm, minimum height=2.05cm, right=1.90cm of r1] (r2)
  {\textbf{role 2}\\ compute\\[3pt]
   {\scriptsize 192 threads, $c$ CTAs per SM}\\
   {\scriptsize 4 math + 1 producer + 1 utility warp}\\
   {\scriptsize $b<G_{\text{relay}}$: relay copiers}};
\node[head, above=2pt of r2] {compute context, $S_{\text{compute}}$ SMs};

\node[ctx, text width=2.95cm, minimum height=2.05cm, right=1.90cm of r2] (r3)
  {\textbf{role 3}\\ finalize\\[3pt]
   {\scriptsize 384 / 256 threads}\\[3pt]
   {\scriptsize combine $k$ slots\\ BF16 store of $y$}};
\node[head, above=2pt of r3] {compute context};

\draw[ar] (r1.east) -- (r2.west);
\node[font=\scriptsize, text=black!65, below=0pt] at ($(r1.east)!0.5!(r2.west)$)
  {programmatic};
\draw[ar] (r2.east) -- (r3.west);
\node[font=\scriptsize, text=black!65, below=0pt] at ($(r2.east)!0.5!(r3.west)$)
  {all compute};

\begin{scope}[yshift=-4.75cm]
  \node[head, anchor=east] at (1.50,2.30) {role 1};
  \node[head, anchor=east] at (1.50,1.85) {shared (fused)};
  \node[head, anchor=east] at (1.50,1.40) {role 2 mainloop};
  \node[head, anchor=east] at (1.50,0.95) {return leg};
  \node[head, anchor=east] at (1.50,0.55) {role 3};

  \node[bar, minimum width=4.70cm] at (1.65,2.30) {scan $\cdot$ bulk pull};
  \node[bar, minimum width=1.35cm, fill=white, draw=black!45, dash pattern=on 2pt off 2pt]
    at (2.35,1.85) {FC1/FC2};
  \fill[black!14] (3.30,1.27) rectangle (11.05,1.53);
  \draw[black!40] (3.30,1.27) rectangle (11.05,1.53);
  \foreach \x in {4.05,4.80,5.55,6.30,7.05,7.80,8.55,9.30,10.05,10.80}
    \draw[black!35, very thin] (\x,1.27) -- (\x,1.53);
  \node[font=\scriptsize] at (7.18,1.40) {GEMM1 + SwiGLU, GEMM2, L1 $\rightarrow$ L2};

  \node[bar, minimum width=7.75cm, fill=white, draw=black!45, dash pattern=on 2pt off 2pt]
    at (5.30,0.95) {coalesced bulk stores / relay copiers};

  \node[bar, minimum width=2.10cm] at (12.10,0.55) {FP32 sum, BF16 $y$};

  \draw[-{Latex[length=1.5mm]}, black!40] (1.60,0.08) -- (14.60,0.08);
  \node[font=\scriptsize, text=black!50, below] at (8.10,0.02) {time};
\end{scope}
\end{tikzpicture}
\caption{(a) The three role kernels and the two Green contexts they run in. (b) One replayed
layer: the phases overlap because no barrier separates them---an expert starts as soon as its
last row lands, and the return leg of one expert runs under the next expert's MMA.}
\label{fig:schedule}
\end{figure}
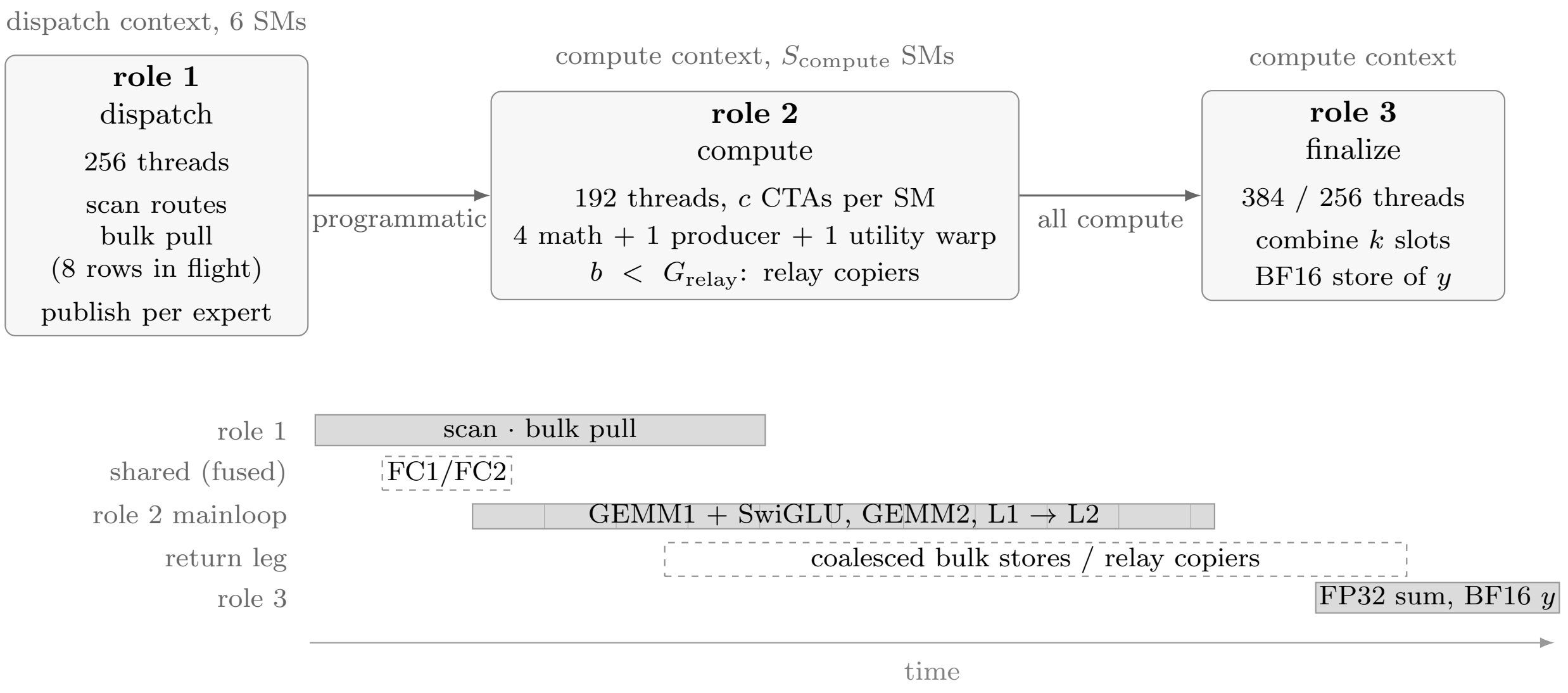

Figure~\ref{fig:schedule}(a) draws the three kernels and the two contexts they occupy, and
panel~\ref{fig:schedule}(b) shows the schedule of one replayed layer. Two properties of the graph
are load-bearing. First, the edge from role 1 to the compute
roles is a \emph{programmatic} dependency: the compute nodes are released once the dispatch
CTAs have taken their seats, not when dispatch finishes, and the actual data dependency is
carried by the per-expert readiness words. Second, all roles keep their synchronization state
in the workspace and derive their barrier phases from monotone issuance counters, so replaying
the graph never re-initializes anything; this is what allows the whole layer to be captured
and replayed a hundred times per timing window.

\subsection{Green-context partition and residency}

The runtime queries the device SM resource, splits it by count with the co-scheduling
alignment $a$ reported by CUDA,
\begin{equation}
S_{\text{compute}} = \Big\lfloor \frac{S_{\text{total}} - S_{\text{dispatch}}}{a}\Big\rfloor a ,
\qquad S_{\text{dispatch}} = 6 \ \text{(default)},
\label{eq:split}
\end{equation}
and creates two green contexts, each with its own stream. It then verifies that the split is
usable ($S_{\text{compute}} \ge 32$, $S_{\text{compute}}$ even, $S_{\text{dispatch}} > 1$,
and the two parts sum to the device count). Both contexts request the balanced workqueue so
that a spinning consumer in the compute context cannot starve the producer.

Residency is checked, not assumed. Before a child graph is instantiated the runtime asks the
driver for the number of active blocks per SM at the role's exact thread and dynamic-SMEM
footprint and rejects the graph if
\begin{equation}
G_{\text{role}} \;>\; c_{\text{role}} \cdot S_{\text{role}},
\label{eq:residency}
\end{equation}
because the role's grid-wide rendezvous would otherwise deadlock. The check covers all three
roles, so neither the shared-expert staging nor the relay arena---both of which raise role 2's
dynamic footprint---can smuggle in a grid that only fits in one of the two modes. It is also
what makes the geometry of \S\ref{sec:geom} authoritative: the heuristic may not propose a
grid the partition cannot hold.

\subsection{Swap-AB block-scaled mainloop}
\label{sec:qmma}

\paragraph{Why swap A and B.} The QMMA has shape $m16n8k32$ and takes $A$ from four 32-bit
registers and $B$ from two. We assign
\begin{center}
$A$: FP4 routed weights, $M{=}64$ weight rows,\qquad
$B$: FP8 activations, $N{=}$ token columns.
\end{center}
This has three consequences. (i) The weight operand is the one with the $M{=}64$ extent, so the
weight TensorMap, the weight SMEM tile and the weight swizzle are the same for every shape;
only the number of $N{=}8$ groups issued against that tile varies with the token count.
(ii) The FP4 operand is loaded by the 4-bit matrix load, which cannot transpose, and a
$K$-major packed weight tile is exactly what a tile-invariant descriptor produces.
(iii) Four math warps each own one $M16$ atom of the $M64$ slice, so the reduction over $K$ is
replicated four times and no warp ever needs another warp's accumulator.

\paragraph{Packed scales with byte-id immediates.} A $K{=}128$ block carries four UE8M0 scale
bytes per lane, i.e.\ one 32-bit word. The mainloop loads that word once per stage with an
ordered SMEM read and then issues four QMMAs whose scale operands are the \emph{same} word
paired with a byte index:

\begin{lstlisting}[caption={The QMMA issue with packed scale words (one of four $K32$ steps).},label={lst:qmma}]
mma.sync.aligned.kind::mxf8f6f4.block_scale.scale_vec::1X.m16n8k32
    .row.col.f32.e2m1.e4m3.f32.ue8m0
    {$0, $1, $2, $3},      // d: 4 x f32 accumulator
    {$4, $5, $6, $7},      // a: FP4 weights, 4 x b32
    {$8, $9},              // b: FP8 activations, 2 x b32
    {$0, $1, $2, $3},      // c
    {$10}, {$11, $12},     // sfa word, byte-id and thread-id
    {$13}, {$14, $15};     // sfb word, byte-id and thread-id
\end{lstlisting}

With $\text{byte\_id} = k_{\text{inner}} \in \{0,1,2,3\}$ the instruction itself selects the
$K32$ sub-block's exponent, so the inner loop contains no shift, no logical-op and no
permute for scales. The alternative---extracting a byte per issue and re-materializing the
SFA/SFB fragments---costs several arithmetic instructions per QMMA in the persistent $K$ loop,
which is precisely the loop that must stay issue-bound. We keep the packed word in a register
across the four issues and reuse it for every $N8$ group of the stage.

\paragraph{Routing the FP4 operand.} Two paths produce the $A$ fragment.

\emph{Expanded path.} A 4D TensorMap stages the weight tile with an FP4$\rightarrow$8-bit
expansion during the bulk copy, so SMEM holds one byte per weight element and the tile is
$8$\,KB per $K128$ step. The warp then loads its fragment with \code{ldsm.b4x16\_p64}
(four matrices, 4-bit unpack, no transpose) and shifts the unpacked E2M1 codes into the bit
position the mixed-input encoding consumes:
\begin{verbatim}
fragment_i8[i] = Int8(fragment_i8[i] << 2);
\end{verbatim}

\emph{Compact path.} Storing the raw nibble tile instead is $4$\,KB per step, half the bytes,
but the fragment must be built by hand: for each of four words the warp issues
\code{ld.shared.u16}, widens it, splits high and low nibbles with \code{prmt}, masks to
$\mathtt{0x0f0f0f0f}$ and shifts left by two.
\begin{verbatim}
ld.shared.u16 h, [address+off]; cvt.u32.u16 packed, h;
shr.u32 high, packed, 4;
prmt.b32 expanded, packed, high, 0x5140;
and.b32 expanded, expanded, 0x0f0f0f0f;
shl.b32 out, expanded, 2;
\end{verbatim}
Because the load bypasses CuTe's pointer swizzle, the compact tile applies SW64 explicitly
($o \oplus ((o \gg 3) \mathbin{\&} \mathtt{0x30})$ for byte offset $o$), and the four sub-words
are fetched with one 16-bit load each at the $0/512/8/520$ byte offsets that the $K$-major
layout implies. The
trade is four scalar SMEM reads per fragment against 4\,KB of SMEM: the compact path is
selected exactly for the tile widths where the freed 4\,KB buys a second co-resident CTA
(\S\ref{sec:geom}).

\paragraph{Scale staging and the $K$ loop.} SFA (weights) and SFB (activations) are staged by
\code{cp.async} into a $512$-byte-per-stage layout (\code{128 * K128 / 32}) and are part of the
same arrival barriers as the data: \code{fullW[s]} completes when the weight TMA \emph{and}
the SFA staging for slot $s$ have landed, \code{fullX[s]} likewise for activations and SFB.
The consumer keeps a compile-time-bounded number of live $N8$ groups, masks the tail with
\code{active\_n\_groups}, and for each group walks $k_{\text{inner}} = 0..3$, so the number of
QMMAs per expert tile is
\begin{equation}
N_{\text{QMMA}} \;=\; \frac{N_{\text{out}}}{8}\;\cdot\;\Big\lceil \frac{r}{\tile}\Big\rceil\;\cdot\; \frac{K}{128}\;\cdot\;4 ,
\label{eq:nqmma}
\end{equation}
with $N_{\text{out}}$ the weight-row extent of the projection ($2I$ for GEMM1, $H$ for GEMM2).

\paragraph{SMEM layout.} Both data operands use the SM120 TMA/ldmatrix-compatible SW128
staging layout built from the warpgroup SMEM atom, so the same descriptor serves the expanded
weight tile, the compact weight tile and the activation tile with different extents.

\subsection{Adaptive tile geometry}
\label{sec:geom}

All geometry is a pure function of the launch shape, evaluated once and cached. From
\eqref{eq:rows} the heuristic computes $r$ and then walks the following decision chain.

\paragraph{Token tile.} When the BF16 handoff path is active, the tile is chosen from four
tiers,
\begin{equation}
\tile \in \{\,32,\ 48,\ 80,\ 96 \,\},\qquad
\tile = \begin{cases}
32, & 16 \le r < 32,\\
48, & 32 \le r < 48,\\
80, & 64 \le r < 96 \ \wedge\ r + \lfloor\sqrt{r}\rfloor \le 80,\\
96, & 64 \le r < 96 \ \wedge\ r + \lfloor\sqrt{r}\rfloor > 80,
\end{cases}
\label{eq:tile}
\end{equation}
Row counts outside those tiers fall back to $\tile = 16$ when $r < 16$ and to $\tile = 64$
otherwise---which is what covers $48 \le r < 64$ and $r \ge 96$ on the handoff path. The
fallback is not free: only the $\{48,80,96\}$ tiers enable compact weights
(\S\ref{sec:qmma}), so a $64$ tile stages full-width weights. The $\sqrt{r}$ slack in
the $80$ tier is the same guard the datacenter kernel uses: uneven routing can push an expert
past its mean row count, and $80$ tiles absorb that without falling back to $96$.

\paragraph{Depth, compactness, handoff.} The pipeline has four stages at $\tile = 16$, where
the tiles are too small to hide latency with only two, and two stages otherwise. Compact
weights are enabled for $\tile \in \{48, 80, 96\}$---exactly the tiers where $4$\,KB decides
co-residency. A global handoff is used for every $\tile \ge 32$ on the BF16 path, and when
$\tile \ge 48$ the handoff is replaced by shared 32-row chunks so the L1 tensor never leaves
the SM. Handoff slots ($s{=}2$ with a global handoff, $s{=}1$ otherwise) let the utility warp
work on one tile while the math warps fill the other.

\paragraph{Return staging mode.} Three booleans decide how a tile's rows leave the SM.
\emph{Batched returns} need the shape to promise at least $32$ rows per local expert
($Bk \ge 32L$) and the workspace to carry the batched route array; below that the operator
falls back to the per-slot route words and the small-tile path. \emph{Mirrored metadata} is
provisioned only when $k < R$, because that is the only case in which a token's activation
scales and route weights are not already local. \emph{Relay staging} then follows: it is on
whenever batched returns are on and the position-major arena fits the relay budget
($R\,B_{\text{cap}}\,k\,2H \le 256$\,MB), unless mirrored metadata is on and no shared
expert is fused---there the direct path already has everything locally and the arena would
only add a copy. For the two shapes of \S\ref{sec:eval} this gives relay staging and the pull
cache for GLM5.2 ($k{=}R{=}8$), mirrored metadata without either for DSV4.1-Flash
($k{=}6<8$), and relay staging for both fused variants, where the shared work hides the copy.

\paragraph{Role-2 SMEM model.} The budget that drives $c$ in \eqref{eq:ctas} is
\begin{equation}
S_{\text{role2}} = \max\big\{\, S_{\text{model}},\ 2S_{\text{shared}} + 16,\ 2\,(16\,\text{KB} + 8)\,\big\},
\label{eq:smem}
\end{equation}
with the mainloop model
\begin{equation}
S_{\text{model}} = s_{\text{st}}\big(P_W + P_X\big) + \tile\cdot 16 + 16
  + B_{\text{bar}} + B_{\text{ret}} + B_{\text{util}} + 512,
\end{equation}
per-stage weight and activation bytes
\begin{equation}
P_W = \frac{64\cdot 128}{2^{c_w}} + 512,\qquad
P_X = \tile\cdot 128 + 512,
\end{equation}
with $c_w = 1$ on the compact-weights path and $0$ otherwise; the $512$ is the per-stage
scale block. Here $s_{\text{st}}$ is the mainloop depth, $B_{\text{bar}}$ the stage-arrival
barriers, the return staging $B_{\text{ret}} = \tile(20 + 128 s)$, which falls to $20\tile + 16$
with a global handoff and to
$20\tile + 128\cdot 32\,s$ once shared 32-row chunks replace the global source tiles, and the
utility overlay $B_{\text{util}} = 32 + 24s + 16s\tile$ for $s$ handoff slots. The second term is
the shared-expert double buffer
\begin{equation}
S_{\text{shared}} = \left\lceil \frac{8192 + 512 + 132\,\tile_s}{1024}\right\rceil \cdot 1024,
\qquad \tile_s \in \{8,16,32,64\},
\end{equation}
i.e.\ one $8$\,KB weight tile, one $512$-byte scale block and $\tile_s$ padded token rows; the
shared GEMM runs two of them and then hands the storage to the routed mainloop. The third term
is the copier's two $16$\,KB slots. These are maxima rather than sums because the three phases
never overlap inside a CTA: the fused and relaying variants add at most the largest of three
footprints instead of their total. The kernel verifies the model at run time: the first thread
of the CTA compares the allocator's reported footprint against this budget with a
trap-on-exceed check, so a compiler change that perturbs the SMEM plan fails loudly instead of
corrupting the handoff.

\paragraph{Reserved return CTAs.} $\ell$ CTAs of the compute grid---the \code{l2\_dedicated}
field of the role configuration---are reserved so that return traffic always has a seat: $\ell = 1$ by default, $\ell = 4$ once $\tile \ge 64$ or a
global handoff is in use ($\ell = 6$ for the unbatched $\tile = 80$ tier, which also needs an
FC1/FC2 scheduling period of slack), and $\ell = 3$ when batched returns are active without
resident copiers and no shared expert is fused. The relay grid is then
\begin{equation}
G_{\text{relay}} = \min\big(R,\ G_{\text{compute}} - \ell\big),
\label{eq:relay}
\end{equation}
and a shape whose compute partition cannot host $\ell + 1$ CTAs is rejected rather than
silently degenerating. The finalize grid is $\min(2S_{\text{compute}},
\lceil B_{\text{bound}} H / 2048\rceil)$, or $\min(B_{\text{cap}}, 4S_{\text{compute}})$ on the
batched path where the epilogue is register-limited ($B_{\text{cap}}$ is the per-rank token
capacity). The dispatch grid is
$S_{\text{dispatch}} \min(4, \lfloor S_{\text{sm}} / S_{\text{role1}}\rfloor)$ with
$S_{\text{role1}} = 512 + 16H + 64$ when the bulk-pull engine is used.

Table~\ref{tab:geom} instantiates the chain for the two evaluation shapes at $B{=}256$. The
interesting contrast is that the models take opposite return paths: GLM5.2 has
$k{=}R{=}8$, so its pull cache engages and the mirrored metadata is unused, while
DSV4.1-Flash has $k{=}6<R$, so its scales and route weights are mirrored, the cache stays
off, and the direct path writes 64-row combine cells with no relay at all---hence
$\ell{=}3$ instead of $4$. Fusing the shared expert turns the relay back on for DSV4.1-Flash
because the shared work hides the copy, which raises $\ell$ to $4$ and makes the two variants
of the second model differ in exactly one scheduling decision.

\begin{table}[t]
\centering
\small
\begin{tabular}{@{}lcc@{}}
\toprule
Geometry & GLM5.2 ($L{=}32,k{=}8$) & DSV4.1-Flash ($L{=}48,k{=}6$) \\
\midrule
$B = 1$: rows/expert, tile, stages & $1$, $16$, $4$ & $1$, $16$, $4$ \\
$B = 256$: rows/expert & $64$ & $32$ \\
$\tile$ / stages & $80$ / $2$ & $48$ / $2$ \\
compact weights & yes & yes \\
handoff: global + shared chunks & yes & yes \\
batched BF16 returns & yes & yes \\
mirrored metadata & no ($k = R$) & yes \\
resident relay copiers & yes & no / yes \\
pull cache & yes & no ($k < R$) \\
$\ell$ (\code{l2\_dedicated}) & $4$ & $3$ / $4$ \\
\bottomrule
\end{tabular}
\caption{Geometry selected at run time for the two models of \S\ref{sec:eval} at $B{=}256$.
Slashes separate the routed operator from the fused shared-expert variant.}
\label{tab:geom}
\end{table}

\subsection{Streaming dispatch}
\label{sec:dispatch}

Role 1 owns several mechanisms beyond the ring protocol inherited from GC-MoE.

\paragraph{Expert-granular publish.} The naive group rendezvous waits for every CTA's share of
eight experts and only then exposes all of them, so the compute partition idles behind the
slowest CTA of the group. Instead, each pulling warp issues a release-add on
\code{expert\_arrive[slot]} after it drains a row, and the thread that completes the expert
writes the count, releases \code{expert\_ready[slot]}, and appends the expert to an
\emph{active} queue that role 2 claims in order. Visibility is transitive by construction: the
completing thread's acquire-release add acquires the whole release sequence of the pullers,
and its own release carries the rows to the claimer's acquire. A separate counter lets the last
expert publish the final queue size, so a consumer never has to guess whether the queue is
drained.

\paragraph{Batched bulk pull.} A fused \code{ld.global} / \code{st.global} pair per $512$-byte
chunk stalls a full PCIe round trip per chunk. Two designs fix this. The engine path issues
\code{cp.async.bulk} peer$\rightarrow$SMEM copies from lane 0 with \emph{two slots per warp} and
one mbarrier per slot; the copy's completion transaction flips the slot's phase, so a warp can
post the next row into the other slot before draining this one, keeping two copies in flight.
The drain then performs only local writes, which keeps the engine dedicated to peer reads. The
fallback path (used when the shape does not admit the engine) batches four chunks in registers
with 16-byte loads before storing them, at a cost of four words per chunk. The batch depth is
set by the shape: $8$ rows in flight when a rank owns fewer than $48$ experts and $1$
otherwise, raised to $4$ when the pull cache is active and lowered to $2$ when batched
returns run without mirrored metadata.

\paragraph{Producer-side pull cache.} When $k \ge R$ a token is routed to several local experts
with high probability, so the same peer row is requested more than once. The cache stages the
row in the local arena on the first request and serves later requests from there. Its state is
\emph{one} 64-bit word per (source rank, source token), claimed by
\begin{verbatim}
atom.acq_rel.gpu.global.cas.b64 entry, [slot], 0, 1;
\end{verbatim}
where $0$ is empty, $1$ means another warp is filling the row, and any larger value is the
local row address itself, published by the draining warp with a release store
(\code{st.release.gpu.global.u64}) once its vector stores have landed. Collapsing flag and
pointer into a single word is what makes a hit waitable: the requester spins on an acquire
load until the entry leaves $1$, then points the bulk engine at that address after an
async-proxy fence. The wait cannot deadlock, because a warp publishes its own outstanding row
before it waits on another warp---the engine keeps one pending row per warp and every
dispatch CTA is resident, so the wait graph stays acyclic. The cache costs $8$ bytes per
(source, token), i.e.\ $8\,R\,B_{\text{cap}}$ bytes in the handoff arena, and is enabled
only when it can be free: the arena must be large enough, $k \ge R$, and the chosen tile must
be $80$, or $48$ when returns are batched---the two geometries where the handoff arena is not
already consumed by return staging.

\paragraph{Packed routes and mirrored metadata.} Two PCIe read streams are removed from the
compute partition. First, in batched mode routes travel as a packed array indexed by
$(\text{peer}, \text{source}, \text{position})$---the same index the return arena uses---instead
of one 32-bit word per $(\text{expert}, \text{source}, \text{slot})$, with a per-expert base that
maps a position back to the source; role 2 therefore needs one array read per hit instead of a
strided peek into the peer's \code{source\_index}. Second, when mirroring is active ($k<R$),
role 1 copies the token's activation scales ($H/128$ words) and its route weights ($k$ words)
into the peer's workspace alongside the route, and publishes the peer epoch only after a system
fence covers both, so the compute partition reads scales and weights locally instead of over
PCIe. The two are alternatives rather than companions: at $k \ge R$ the deduplicating cache
already removes the repeated activation reads, and the scale/weight mirror is not allocated at
all.

\paragraph{Tiny batches.} For $B \le 2$ the operator uses a direct scan: one block polls all
peers over the fabric and opens a local gate, so the remaining blocks spin on local memory
instead of re-issuing sys-scope acquires, and the expert grid is clamped to the number of local
experts.

\subsection{Return path: full-width BF16 and relay copiers}
\label{sec:return}

\paragraph{Motivation.} The down-projection result is BF16, $2H$ bytes per hit: twice the
forward activation and, on PCIe, the largest term of \eqref{eq:volume}. An earlier version of
this operator coded each element in 12 bits to shrink that payload. The current design drops
the codec. A $96$-byte row breaks the $128$-byte row granularity that both the combine cells
and the bulk copies want, the encode/decode arithmetic ran on the same utility warp that
stages the row, and every receiver would have had to decode before its FP32 accumulate. It is
cheaper to keep the row full width and make the copy itself the object of optimization.

\paragraph{Direct path: coalesced bulk stores.} The utility warp owns the L2 epilogue. It
first materializes, for every valid row of the tile, the 64-bit destination address inside the
\emph{combine arena}: 64-row cells of $128$-byte rows, one cell per column block and N-tile,
so that a token's output for one column block is contiguous. Consecutive positions that share
a source and fall inside the same cell therefore differ by exactly $128$ bytes. The warp
ballots row-group boundaries and issues one \code{cp.async.bulk} per run of consecutive
destinations, so a burst of positions becomes a single bulk store instead of one store per
row. One copy group is kept in flight (\code{commit\_group}, then \code{wait\_group(1)}),
which hides the previous round's store latency without ever draining the queue.

\paragraph{Relay path: position-major staging, resident copiers.} Bulk stores issued from the
compute partition still cost a PCIe write from a CTA that could be doing arithmetic. The relay
path moves them out. The utility warp parks each finished row into a local arena at
\code{(row\_index * colblocks + ntile) * 128}, where
\code{row\_index = source * B\_cap * k + pos}: the arena is position-major, so all rows of one
(expert, source) pair are contiguous no matter which N-tile produced them. Role 2 then splits
its grid: the first $G_{\text{relay}}$ blocks \emph{are} the copiers (\eqref{eq:relay}) and
the rest run the MMA, so the copiers cost co-residency but no extra SMEM---their two $16$\,KB
slots alias the mutually exclusive MMA workspace. Each copier walks, for one peer, the experts
in the order they become active, waits for the expert's L2 completion count
(\code{relay\_done}), and streams \code{count * colblocks * 128} bytes
global$\to$shared$\to$global in $16$\,KB chunks, retiring the older send before it overwrites
the source slot. Because it follows the FC2 active queue rather than the expert id, a slow
expert cannot hold up output that is already complete.

\begin{figure}[t]
\centering
\begin{tikzpicture}[font=\footnotesize,
  blk/.style={draw=black!45, rounded corners=2pt, fill=black!3, align=center, inner sep=3pt},
  ar/.style={-{Latex[length=1.7mm]}, semithick, black!60},
  note/.style={font=\scriptsize, text=black!70, align=center}]

\node[blk, text width=2.5cm] (epi) at (0,1.95) {L2 epilogue\\ (utility warp)};
\node[blk, text width=3.0cm] (cell) at (4.85,1.95) {combine arena: $64\times64$ cells of
  128-byte rows};
\draw[ar] (epi) -- node[above, font=\scriptsize, text=black!70] {coalesce} (cell);
\node[note, anchor=north] at (4.85,1.36)
  {consecutive destinations $\Rightarrow$ one \texttt{cp.async.bulk} per run};

\node[blk, text width=2.8cm, anchor=west] (comb) at (11.85,1.95)
  {role 3: FP32 sum of $k$ slots\\ then BF16 $y$};
\draw[ar] (cell) -- (comb);

\node[blk, text width=3.2cm] (arena) at (4.85,0.30)
  {position-major arena\\ $(\text{row\_index}\cdot C + n)\cdot128$};
\node[blk, text width=2.1cm] (cop) at (9.55,0.30) {copier CTA\\ $b<G_{\text{relay}}$};
\node[blk, text width=1.5cm] (peer) at (13.30,0.30) {peer};
\draw[ar] (arena) -- node[above, font=\scriptsize, text=black!70] {16\,KB} (cop);
\draw[ar] (cop) -- node[above, font=\scriptsize, text=black!70] {bulk} (peer);
\node[note, anchor=north] at (9.55,-0.30) {double-buffered; follows\\ the FC2 active queue};
\end{tikzpicture}
\caption{The two return paths. Top: the direct path writes each row where the combine will read it
and merges runs of consecutive destinations into one bulk store. Bottom: the relay path parks rows
position-major and hands them to a resident copier CTA.}
\label{fig:return}
\end{figure}
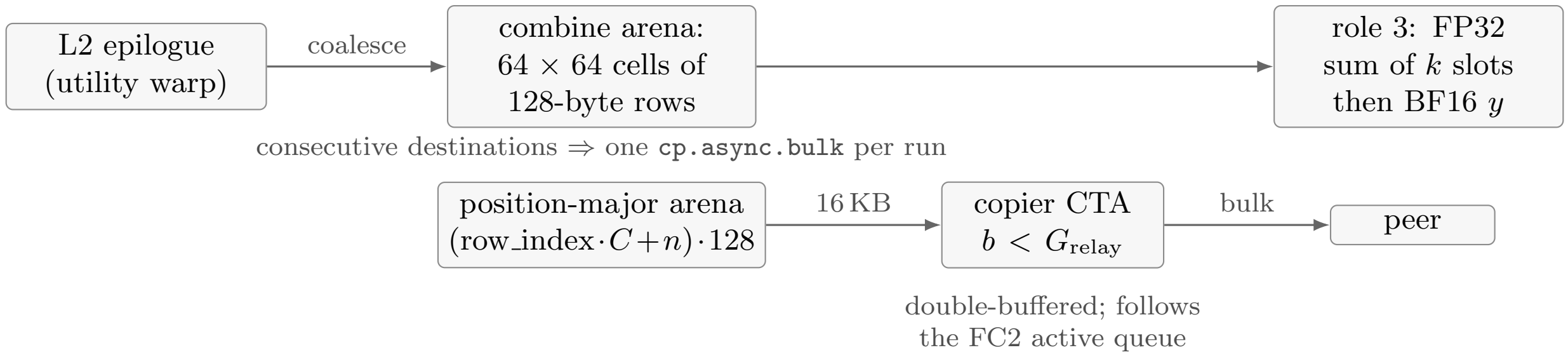

\paragraph{Combine.} Role 3 then accumulates the $k$ slots in FP32 and writes $y$ once, in
BF16; Figure~\ref{fig:return} contrasts the two ways a row reaches it. The addressing is the only
thing that differs between the two paths---the direct path
reads $64\times64$ cells, the relay path reads position-major rows---so the same finalize
kernel is compiled with one boolean. A route that lost its arena slot to capacity is marked
with an out-of-range position and poisons that token's sum rather than silently reading a
neighbouring row.

\paragraph{Measured cost.} The return leg is deliberately uncompressed, so its wire volume is
the $2H$ of \eqref{eq:volume}; what the operator buys instead is transaction efficiency and
SM independence. At $B{=}128$ (GLM5.2) the returned payload moves at $24.1$\,GB/s per GPU
against $12.0$ for the CuTe DSL operator and $18.8$ for Cake, and at $B{=}256$ all three
converge near $30$\,GB/s, where the leg stops separating them and the weight stream takes
over.

\subsection{Fused shared experts inside the compute role}
\label{sec:se}

A shared expert runs on every token, so its GEMMs are token-parallel rather than
expert-parallel. Instead of two further kernels, FC1 and FC2 are compiled into role 2 and run
before the routed mainloop against the same dynamic SMEM allocation. Only the first $128$
threads (four warps) take part; the two remaining warps idle, which is free because the shared
work completes before the mainloop starts.

The shared GEMM is a small mainloop of its own: a double-buffered $K{=}128$ pipeline of one
$64\times128$ weight tile ($8$\,KB), a $512$-byte scale block and $\tile_s$ padded token rows
($\tile_s \in \{8,16,32,64\}$, chosen from the token count), with two mbarriers and a
\code{barrier.sync 2, 128} that names the participating warp count rather than the block. FC1
stages the token-major activation rows, applies BF16 SwiGLU with the caller's clamp---the clamp
bound is itself rounded to BF16 so that both branches agree---and requantizes to MXFP8 with
UE8M0 scales; FC2 multiplies by the shared down projection and writes \code{shared\_y}.
Shared weights are FP8 with a $K$-group of $32$ or $128$ and must arrive as the packed int32
scale payload, which is the only form the shared mainloop accepts.

Ordering is carried by counters, not by graph edges. Tiles are indexed so that all FC1 output
tiles of one token tile are adjacent; a worker release-adds that token tile's counter when its
FC1 tile finishes, and an FC2 tile waits until the counter reaches the number of FC1 output
tiles in the group. Every worker finishes its own FC1 sweep before entering the FC2 sweep and
all role-2 CTAs are resident, so the wait is always satisfiable. At the end the block
synchronizes, the shared mbarriers are invalidated, and the routed mainloop takes over the same
storage: the shared path adds neither a role, nor a graph edge, nor persistent SMEM.

On the finalize side, role 3 adds \code{shared\_y} into the same FP32 accumulate that folds
the $k$ routed slots, so the shared expert costs no extra pass over $y$, and block 0 of the
finalize clears the FC1 counters from both finalize paths before the workspace is reused. That
last detail is what keeps a captured graph replayable without a reset pass between
layers---the reason the graph could be replayed a hundred times per timing window.

\subsection{Warp specialization and the epilogue}

Role 2 has 192 threads: four math warps (one $M16$ atom each), one TMA producer warp, and one
utility warp. The producer owns separate W and X barrier sets, so a stalled weight descriptor
never blocks activation staging, and the scale \code{cp.async} rides the same arrival as its
data. The utility warp runs the L1 epilogue (route-weight multiply, SwiGLU, requantization) and
the L2 return staging on alternating handoff slots; its FP32 FC1 handoff overlays the return
staging storage, which is how a $\tile{=}96$ configuration still fits the SMEM model of
\eqref{eq:smem}. Route sources and positions for the whole tile are prefetched into registers
during the MMA so that the epilogue never issues a dependent global read.

\paragraph{L1 epilogue.} For each 16-row group the utility warp multiplies the gate and up
tiles by the route weight, evaluates the clamped SwiGLU in BF16, and requantizes to MXFP8 with
one UE8M0 exponent per 32-element block, writing activations and scales into the L2 arena in
the layout the second mainloop stages from. The staging that follows is either the coalesced
bulk store or the relay park of \S\ref{sec:return}; both are chosen at compile time from the
same shape decision, so the warp has no branch in its loop.

\section{Implementation}

\subsection{PTX rewriting and occupancy control}

The role kernels are written in CuTe DSL and compiled to \code{sm\_120a} cubins at run time,
keyed by a compiler identity (release string, target, ptxas options). Four rewrites happen
between the DSL's PTX and ptxas:

\paragraph{PTX version.} CuTe emits a \code{.version 9.4} header for these kernels. Under the
CUDA 13.2 toolchain we rewrite it to \code{.version 9.2}: the SM120 instruction stream we emit
is legal under 9.2, and ptxas verifies every instruction against the target architecture
anyway. The rewrite is applied only when the recorded toolchain release is 13.2, so a newer
toolchain is left untouched.

\paragraph{Register ceiling.} The warp-specialized producer adds enough live values to tip the
role-2 CTA past the register file once the allocator's 16-wide per-thread rounding lands: the
kernels compile to just under two hundred registers per thread, the allocator rounds the
number up, and at $192$ threads two CTAs no longer fit the 65,536-register
file---both the block limit and the driver's L1/shared carveout then collapse to a single CTA
per SM, a $\sim$1.5$\times$ slower sweep. The runtime therefore rewrites CuTe's
minimum-occupancy directive from
\code{.minnctapersm 1} to the co-residency the SMEM budget intends, and pins a 16-aligned
ceiling $R_{\text{cap}}$ from \eqref{eq:ctas}: 160 registers per thread at two CTAs per SM,
80 at four, and no ceiling when a single CTA co-resides.

\paragraph{Finalize occupancy.} The 384-thread finalize kernel is emitted with
\code{.minnctapersm 4}. The scheduler only ever asks for two resident finalize CTAs per SM, so
the directive is rewritten to 2; otherwise the driver would shrink the carveout for a CTA count
that is never requested.

\paragraph{TensorMaps by value.} Role 2 needs six descriptors---$W_1$, $X_1$, $W_2$, $X_2$ and
the two shared weight maps---and a descriptor that lives in global memory is fetched into the
tensormap cache on first use. That fetch is exactly what a replayed graph cannot afford: it is
a miss inside the mainloop's first stage, and it is invisible to the graph. The host therefore
encodes all six into $128$-byte blobs and passes them in the argument vector, and the PTX
rewrite turns the four routed map formals (8--11) and the two shared ones (15--16) into
\code{.param .align 128 .b8 name[128]} and their single \code{ld.param.u64} into
\code{cvta.param::entry.u64}, so each TMA instruction addresses the parameter space directly.
The rewrite asserts that every map is loaded exactly once and used exactly twice, which is a
stronger check than counting formals: it fails if CuTe ever reuses a descriptor for a second
purpose. Roles 1 and 3 keep the plain pointer ABI, and role 2 drops the shared pair when no
shared expert is fused.

Each rewrite asserts that it matched exactly once. In addition, every role's PTX is validated
before assembly: the entry must have exactly 20 formals---seventeen pointers and three scalars,
in that order---and the role-2 PTX must still contain both
\code{mma.sync.aligned.kind::mxf8f6f4} and \code{cp.async.bulk.tensor}.
This validation is what keeps the hand-written inline assembly of \S\ref{sec:qmma} honest
across compiler upgrades: if a CuTe version changes the launch ABI or drops the block-scaled
MMA, the build fails with a message naming the lost feature instead of producing a slow or
wrong kernel.

\subsection{Graph construction and caching}

Role cubins are loaded with \code{cuLibraryLoadData}. A CUDA Graph kernel node created through
the driver API cannot carry a \code{CUfunction} for a library kernel, so the runtime fills the
node's \code{kern} and an explicit \code{ctx} (the role's green context, or the caller's primary
context when primary-context roles are requested) and leaves \code{func} null. The dynamic SMEM
size is set per node as an attribute before the node is added.

\paragraph{Descriptors.} Six 2D/4D TensorMaps are encoded on the host---$W_1$, $X_1$, $W_2$,
$X_2$ and, when a shared expert is fused, its two weight maps---as $128$-byte descriptors held
in \code{c\_uint8[128]} blobs. They are cached by the pointer set they embed (workspace,
weight payloads, tile width, compactness and the shared weights) and the cache pins every
tensor it references, so a recycled allocation produces a new descriptor rather than a stale
one. Only role 2 receives them, and it receives them by value rather than through the argument
pointers.

\paragraph{Child-graph cache.} A child graph is identified by a SHA-256 over the per-role arg
bytes, the per-role cubin hashes, and the (smem, threads, grid) triples. Because the TensorMap
blobs are part of the arg bytes, a new workspace or a recycled pointer produces a new child
rather than a stale descriptor. When the caller's stream is already capturing, the child is
attached as a graph node and the capture dependencies are rewritten to it; otherwise it is
instantiated and launched directly. This is what lets a whole GC-MoE layer be nested inside a
larger decode graph and replayed without rebuild.

\subsection{Workspace and synchronization state}

The symmetric workspace is one flat allocation viewed as typed tensors by an explicit layout:
a header, the peer pointer table, per-expert readiness and count arrays, the L1 completion
array, the active-expert queue, the combine-block bitmap, counters and positions, the packed
activation and scale regions, the route metadata, the L1 and L2 activation arenas, the combine
arena, per-expert bases and counts, the handoff arena, and---when enabled---the batched route
arrays (\code{batched\_routes\_out}, \code{batched\_routes\_in}), the mirrored metadata
(\code{batched\_scales\_in}, \code{batched\_weights\_in}), the relay arena
(\code{relay\_output} with one completion word per expert) and the shared-expert regions
(\code{shared\_fc1\_done}, \code{shared\_l2\_acts}, \code{shared\_l2\_acts\_sf},
\code{shared\_y}). Shared storage is appended rather than interleaved, so every routed offset
is identical with and without a fused expert and the two variants share one workspace layout.
Two layout decisions matter for latency.

The combine arena is addressed as 64-row \emph{cells} of 128-byte rows, one cell per column
block and N-tile, so a column block of a token's output is contiguous in one destination and a
returning row can be written with one bulk store. When the relay path is selected the same
arena is instead addressed position-major---one row per (source, position) pair---because that
is the only order in which a contiguous run of rows also belongs to one expert on one peer.
And there is \emph{no trailing cross-rank
barrier}: the receive counters are double-buffered, so layer $n+1$ can start publishing before
layer $n$'s counters are retired, and every wait is an equality test on an epoch word.
Equality remains safe at 32-bit wraparound because all ranks execute the calls in the same order
and the tail rendezvous prevents skipping a still-needed epoch.

\subsection{Numerics}

The K=128 scale words are consumed as UE8M0 exponents, and the reciprocal scale is built as an
exponent subtraction, so requantization costs no division. SwiGLU uses a clamped
$\exp_2$-based sigmoid with the caller's clamp ($10$ in our experiments); the clamp bound, the
gate branch and the up branch are each rounded to BF16 before the multiply, and the epilogue
rounds to BF16 with round-to-nearest-even. Under fast math the sigmoid's reciprocal is
\code{rcp}; without it the division is issued with \code{NEAREST\_EVEN} rounding rather than
the compiler default. The return leg now carries BF16 unchanged (\S\ref{sec:return}), so the
only approximation left in the layer is the input quantization itself.

\section{Evaluation}
\label{sec:eval}

\subsection{Setup}

\begin{table}[t]
\centering
\small
\begin{tabular}{@{}ll@{}}
\toprule
Item & Value \\
\midrule
Hardware & 8 $\times$ NVIDIA RTX PRO 5000 72\,GB (SM120), PCIe \\
Software & GC MoE 0.6.0, FlashInfer 0.7.3, ptxas 13.2.78 \\
Precision & FP8 (E4M3) activations, MXFP4 routed weights \\
Baselines & FlashInfer CuTe DSL W4A8 MegaMoE; Cake MegaMoE (\code{all\_lsa}) \\
Routing & same pre-quantized inputs, weights and routes for all operators \\
Capacity & 256 tokens per rank (largest $B$); SwiGLU clamp 10 \\
Timing & graph warmup 10 windows, 30 timed windows, 100 replays each \\
Statistic & max over ranks of the per-rank median, microseconds \\
Green Context & default split (6 dispatch SMs) \\
\bottomrule
\end{tabular}
\caption{Evaluation environment.}
\label{tab:env}
\end{table}

\begin{table}[t]
\centering
\small
\begin{tabular}{@{}lrrrr@{}}
\toprule
Model & $H$ & $I$ & routed experts & top-$k$ \\
\midrule
GLM5.2 & 6144 & 2048 & 256 & 8 \\
DSV4.1-Flash & 5120 & 2304 & 384 & 6 \\
\bottomrule
\end{tabular}
\caption{Model shapes. The shared-expert runs additionally fuse one shared expert with
$K$-group 128 (GLM5.2) and 32 (DSV4.1-Flash).}
\label{tab:models}
\end{table}

Table~\ref{tab:env} lists the environment and Table~\ref{tab:models} the two model shapes.
All operators use the same pre-quantized activations, weights and routes, so the comparison is of
scheduling and arithmetic, not of accuracy. Latency is the maximum over the eight ranks of each
rank's median, which removes cross-rank skew from the report without hiding the slowest rank.

\subsection{Routed decode latency}

\begin{figure}[t]
\centering
\includegraphics[width=0.86\linewidth]{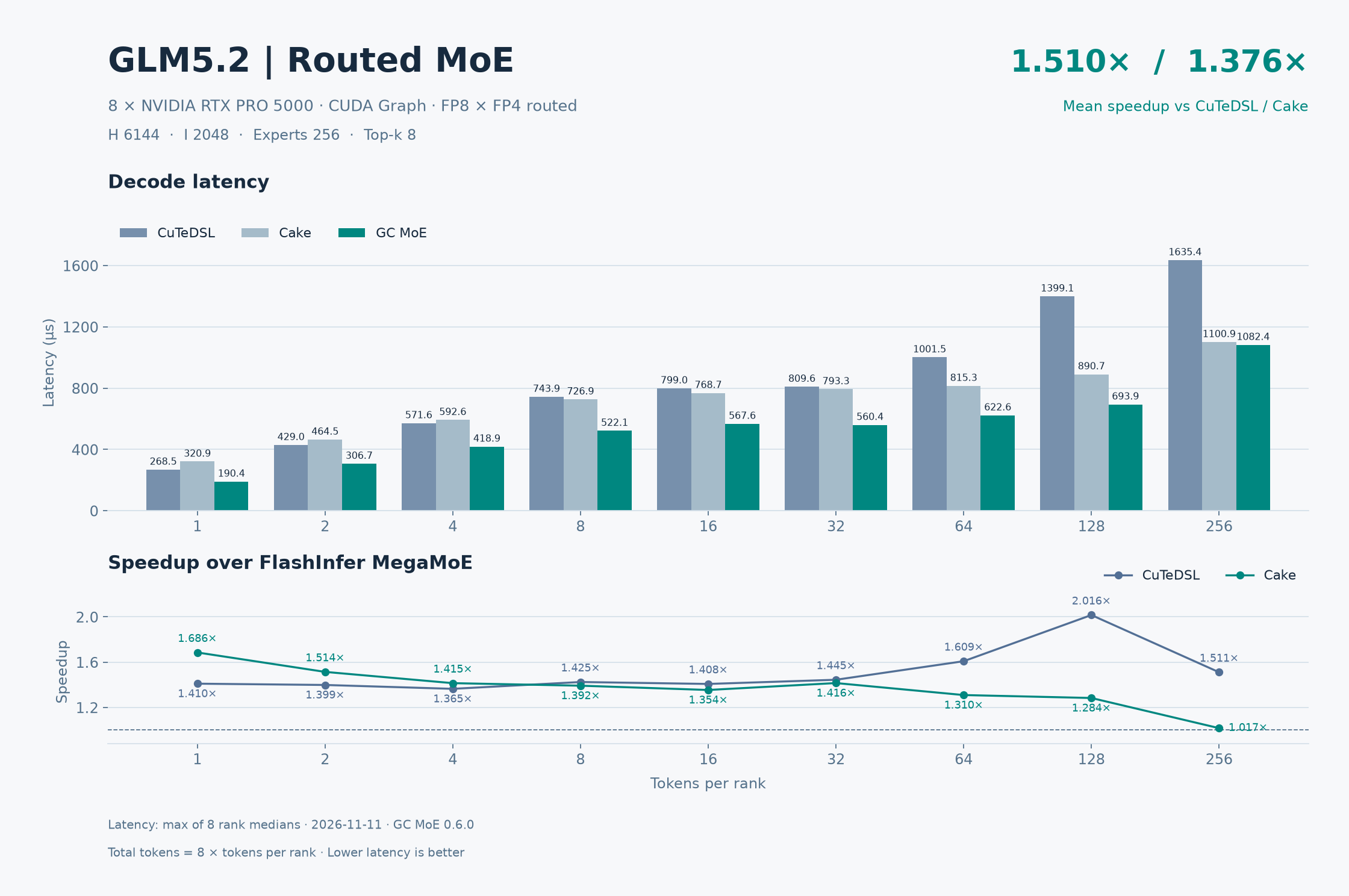}
\caption{GLM5.2: routed decode latency and per-point speedup over the two baselines.}
\label{fig:glm}
\end{figure}

\begin{figure}[t]
\centering
\includegraphics[width=0.86\linewidth]{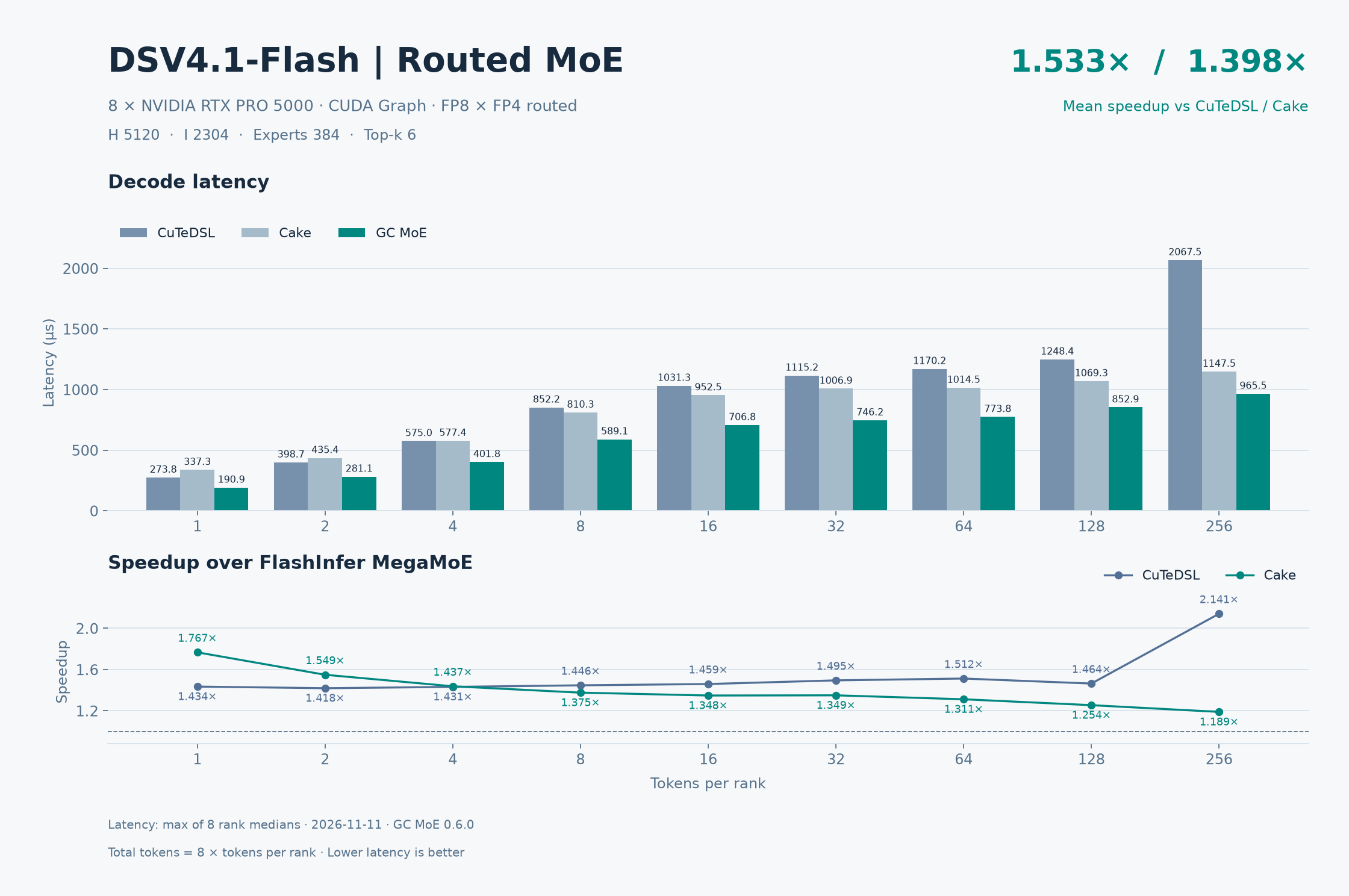}
\caption{DSV4.1-Flash: routed decode latency and per-point speedup over the two baselines.}
\label{fig:dsv}
\end{figure}

Tables~\ref{tab:glm} and \ref{tab:dsv} show the routed operator; Figures~\ref{fig:glm}
and~\ref{fig:dsv} plot the same data. It is faster than both baselines at all 18 points,
averaging $1.522\times$ over CuTe DSL (mean latency reduction $33.34\%$) and $1.387\times$
over Cake ($26.84\%$).

\begin{table}[t]
\centering
\small
\setlength{\tabcolsep}{4pt}
\begin{tabular}{@{}rrrrrrr@{}}
\toprule
$B$ & tokens & CuTe DSL & Cake & \textbf{GC} & DSL/GC & Cake/GC \\
\midrule
1 & 8 & 268.47 & 320.95 & \textbf{190.39} & 1.410 & 1.686 \\
2 & 16 & 429.01 & 464.45 & \textbf{306.73} & 1.399 & 1.514 \\
4 & 32 & 571.58 & 592.61 & \textbf{418.89} & 1.365 & 1.415 \\
8 & 64 & 743.89 & 726.91 & \textbf{522.10} & 1.425 & 1.392 \\
16 & 128 & 798.96 & 768.70 & \textbf{567.56} & 1.408 & 1.354 \\
32 & 256 & 809.59 & 793.29 & \textbf{560.41} & 1.445 & 1.416 \\
64 & 512 & 1001.45 & 815.31 & \textbf{622.57} & 1.609 & 1.310 \\
128 & 1024 & 1399.12 & 890.66 & \textbf{693.87} & 2.016 & 1.284 \\
256 & 2048 & 1635.44 & 1100.85 & \textbf{1082.38} & 1.511 & 1.017 \\
\midrule
mean & & & & & \textbf{1.510} & \textbf{1.376} \\
\bottomrule
\end{tabular}
\caption{GLM5.2 routed decode latency ($\mu$s, lower is better) and per-point speedups.}
\label{tab:glm}
\end{table}

\begin{table}[t]
\centering
\small
\setlength{\tabcolsep}{4pt}
\begin{tabular}{@{}rrrrrrr@{}}
\toprule
$B$ & tokens & CuTe DSL & Cake & \textbf{GC} & DSL/GC & Cake/GC \\
\midrule
1 & 8 & 273.78 & 337.25 & \textbf{190.89} & 1.434 & 1.767 \\
2 & 16 & 398.74 & 435.42 & \textbf{281.11} & 1.418 & 1.549 \\
4 & 32 & 575.03 & 577.39 & \textbf{401.79} & 1.431 & 1.437 \\
8 & 64 & 852.16 & 810.27 & \textbf{589.13} & 1.446 & 1.375 \\
16 & 128 & 1031.32 & 952.48 & \textbf{706.83} & 1.459 & 1.348 \\
32 & 256 & 1115.22 & 1006.87 & \textbf{746.20} & 1.495 & 1.349 \\
64 & 512 & 1170.20 & 1014.47 & \textbf{773.81} & 1.512 & 1.311 \\
128 & 1024 & 1248.37 & 1069.32 & \textbf{852.89} & 1.464 & 1.254 \\
256 & 2048 & 2067.47 & 1147.47 & \textbf{965.46} & 2.141 & 1.189 \\
\midrule
mean & & & & & \textbf{1.533} & \textbf{1.398} \\
\bottomrule
\end{tabular}
\caption{DSV4.1-Flash routed decode latency ($\mu$s) and per-point speedups.}
\label{tab:dsv}
\end{table}

The shape dependence is informative. At $B{=}1$ the layer is control bound: $190\,\mu$s of
latency for eight tokens, where a single dispatch handshake costs tens of microseconds. The
tiny-batch path (direct scan, $\tile{=}16$ with four stages) is what buys the
$1.69$--$1.77\times$ lead over Cake there. The mid range is the flattest part of the curve
($1.35$--$1.44\times$): all three operators then have enough work to stream the same weights,
and no scheduling trick is worth much. Beyond it the two baselines diverge. CuTe DSL degrades
super-linearly---$1001 \to 1635\,\mu$s for GLM5.2 and $1170 \to 2067\,\mu$s for DSV4.1-Flash
from $B{=}64$ to $B{=}256$---where ours grows by $460$ and $192\,\mu$s, because the term that
scales with the token count is the return leg and we have moved it onto copy engines. Cake
instead degrades gracefully, so our margin over it narrows toward full capacity
($1.284 \to 1.017\times$ for GLM5.2, $1.254 \to 1.189\times$ for DSV4.1-Flash between
$B{=}128$ and $B{=}256$).

The $B{=}256$ point of GLM5.2 is the one place where the two operators are effectively tied,
and it is worth being precise about why. At that size the routed weight working set no longer
fits and both operators stream it at about the same rate ($593$ vs $583$\,GB/s per GPU,
Table~\ref{tab:bw}), so the layer becomes HBM bound, the return leg stops being the
differentiator, and $18\,\mu$s is all that separates them. That point is also where our own
design gives up the most: the copier CTAs are taken from the MMA grid, which is exactly the
resource the weight stream needs. We accept the trade because the same mechanism carries
$B{=}64\ldots128$, where every point is $1.28$--$1.31\times$ ahead of Cake while the
uncompressed return is already moving $24.1$\,GB/s. DSV4.1-Flash never pays it: with
$k{=}6<R$ its returns take the direct path, no CTA is given up, and the margin at $B{=}256$
stays at $1.189\times$.

\subsection{Fused shared experts}

\begin{figure}[t]
\centering
\includegraphics[width=0.86\linewidth]{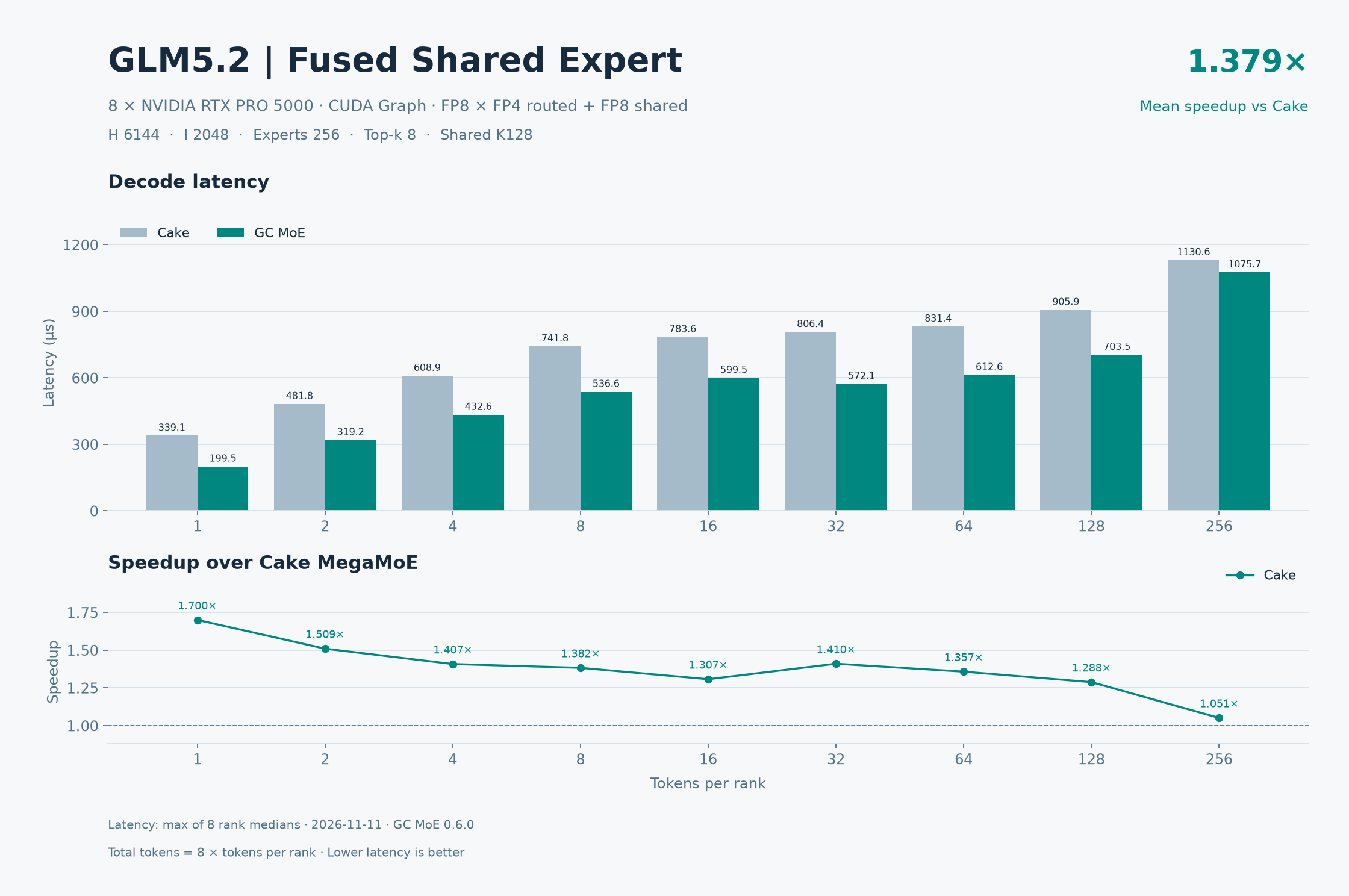}
\caption{GLM5.2 fused shared expert: decode latency and per-point speedup over Cake.}
\label{fig:seglm}
\end{figure}

\begin{figure}[t]
\centering
\includegraphics[width=0.86\linewidth]{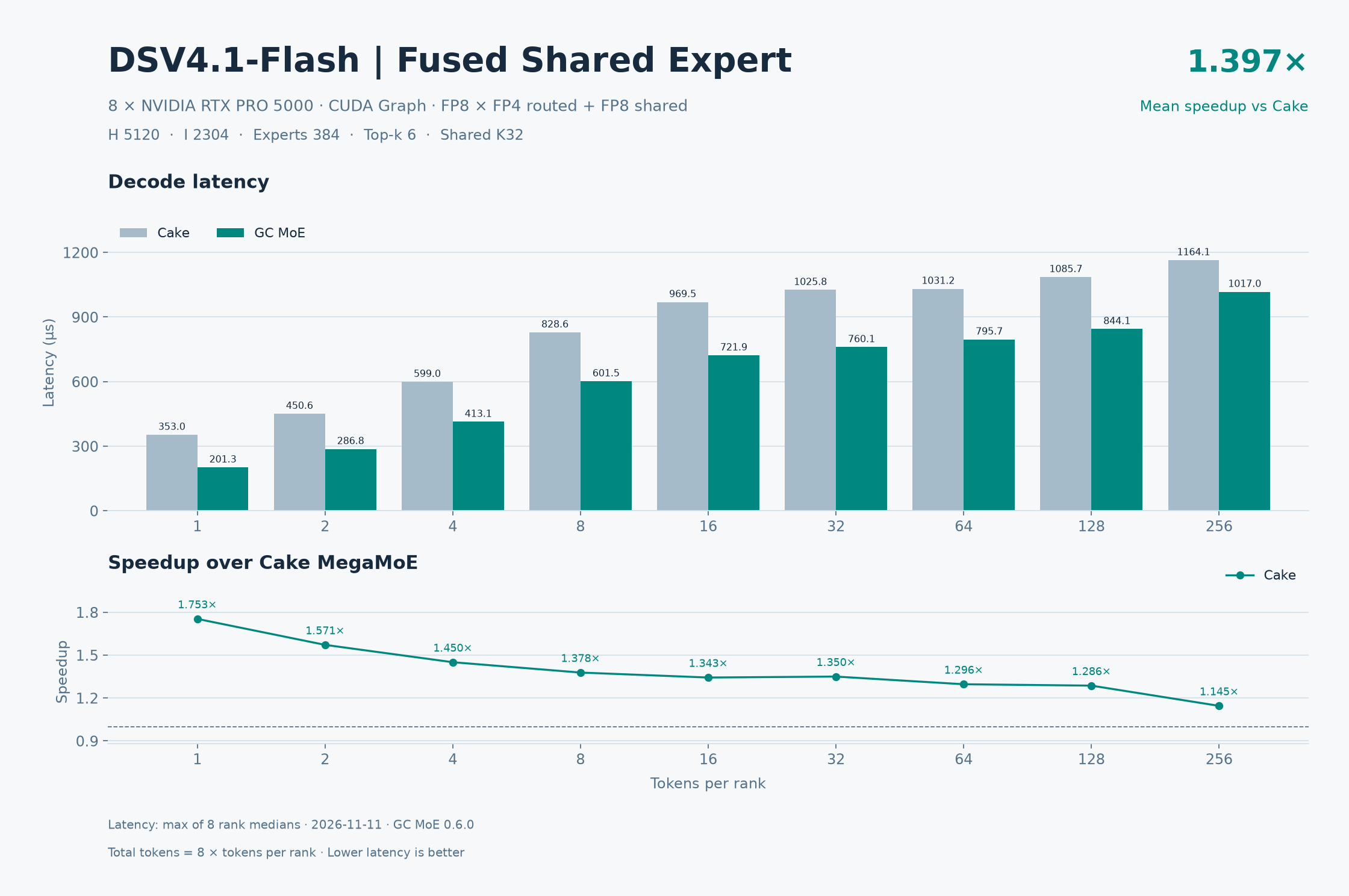}
\caption{DSV4.1-Flash fused shared expert: decode latency and per-point speedup over Cake.}
\label{fig:dsvse}
\end{figure}

Table~\ref{tab:se} reports the fused variant, which additionally computes one shared expert
inside the same role; Figures~\ref{fig:seglm} and~\ref{fig:dsvse} plot the per-point detail.
Average speedup over Cake is $1.388\times$ (18/18, mean latency
reduction $26.92\%$). The small-batch points are the strongest ($1.70$--$1.75\times$): there
the shared expert is one $8$-token tile per CTA, and its cost is a fraction of the dispatch
handshake it overlaps. The mid range sits at $1.29$--$1.45\times$ and the two $B{=}256$ points
fall to $1.05$--$1.15\times$, again where the weight stream saturates HBM.

\begin{table}[t]
\centering
\small
\setlength{\tabcolsep}{4pt}
\begin{tabular}{@{}rrrr|rrr@{}}
\toprule
& \multicolumn{3}{c|}{GLM5.2} & \multicolumn{3}{c}{DSV4.1-Flash} \\
$B$ & Cake & \textbf{GC} & ratio & Cake & \textbf{GC} & ratio \\
\midrule
1 & 339.10 & \textbf{199.51} & 1.700 & 352.95 & \textbf{201.33} & 1.753 \\
2 & 481.78 & \textbf{319.19} & 1.509 & 450.63 & \textbf{286.77} & 1.571 \\
4 & 608.91 & \textbf{432.63} & 1.407 & 598.97 & \textbf{413.08} & 1.450 \\
8 & 741.81 & \textbf{536.62} & 1.382 & 828.64 & \textbf{601.50} & 1.378 \\
16 & 783.59 & \textbf{599.50} & 1.307 & 969.48 & \textbf{721.93} & 1.343 \\
32 & 806.40 & \textbf{572.08} & 1.410 & 1025.84 & \textbf{760.12} & 1.350 \\
64 & 831.42 & \textbf{612.61} & 1.357 & 1031.16 & \textbf{795.73} & 1.296 \\
128 & 905.94 & \textbf{703.50} & 1.288 & 1085.70 & \textbf{844.12} & 1.286 \\
256 & 1130.61 & \textbf{1075.66} & 1.051 & 1164.13 & \textbf{1017.03} & 1.145 \\
\midrule
mean & & & \textbf{1.379} & & & \textbf{1.397} \\
\bottomrule
\end{tabular}
\caption{Fused shared-expert decode latency ($\mu$s) versus Cake, which also fuses one shared
expert.}
\label{tab:se}
\end{table}

\subsection{Effective bandwidth and throughput}

Table~\ref{tab:bw} derives three rates from each operator's own latency: the rate at which the
quantized weight working set is streamed from HBM (one FP4 payload plus scales per hit expert,
plus local FP8 shared weights on non-empty ranks), the effective inter-GPU payload rate of the
exchange (remote routes $\times\, (H + H/32 + 2H)$, charging the same uncompressed return
to all three operators so the comparison is apples to apples), and the effective matrix
throughput ($6\cdot\text{assignments}\cdot H\cdot I$).

\begin{table}[t]
\centering
\small
\setlength{\tabcolsep}{4pt}
\begin{tabular}{@{}lrrr|rrr|rrr@{}}
\toprule
& \multicolumn{3}{c|}{HBM GB/s / GPU} & \multicolumn{3}{c|}{payload GB/s / GPU} & \multicolumn{3}{c}{TFLOPS / GPU} \\
Model & DSL & Cake & \textbf{GC} & DSL & Cake & \textbf{GC} & DSL & Cake & \textbf{GC} \\
\midrule
GLM5.2 $B{=}1$   & 550.9 & 460.8 & \textbf{776.8} & 0.49 & 0.41 & \textbf{0.70} & 2.25 & 1.88 & \textbf{3.17} \\
GLM5.2 $B{=}32$  & 792.7 & 809.0 & \textbf{1145.1} & 5.16 & 5.27 & \textbf{7.46} & 23.9 & 24.4 & \textbf{34.5} \\
GLM5.2 $B{=}128$ & 458.7 & 720.5 & \textbf{924.9} & 12.0 & 18.8 & \textbf{24.1} & 55.3 & 86.8 & \textbf{111.4} \\
GLM5.2 $B{=}256$ & 392.4 & 582.9 & \textbf{592.9} & 20.5 & 30.4 & \textbf{30.9} & 94.5 & 140.5 & \textbf{142.9} \\
DSV4.1 $B{=}1$   & 412.0 & 334.5 & \textbf{590.9} & 0.28 & 0.23 & \textbf{0.41} & 1.55 & 1.26 & \textbf{2.22} \\
DSV4.1 $B{=}32$  & 792.3 & 877.6 & \textbf{1184.2} & 2.37 & 2.62 & \textbf{3.54} & 12.2 & 13.5 & \textbf{18.2} \\
DSV4.1 $B{=}128$ & 722.9 & 843.9 & \textbf{1058.1} & 8.39 & 9.79 & \textbf{12.3} & 43.5 & 50.8 & \textbf{63.7} \\
DSV4.1 $B{=}256$ & 436.5 & 786.5 & \textbf{934.7} & 10.1 & 18.2 & \textbf{21.6} & 52.6 & 94.7 & \textbf{112.6} \\
\bottomrule
\end{tabular}
\caption{Effective per-GPU rates derived from measured latency (selected points).}
\label{tab:bw}
\end{table}

Three observations follow. First, the weight stream is the binding constraint once the batch is
large: our best HBM rate is $1145$\,GB/s per GPU at $B{=}32$ (GLM5.2), $1.42\times$ the Cake and
$1.45\times$ the CuTe DSL figure, which is only reachable if dispatch and return never stall the
weight pipeline. Second, the payload rate at $B{=}128$, $24.1$\,GB/s per GPU, is twice the CuTe
DSL rate---the deduplicated reads, the batched routes and the copiers all show up here at
once, and none of them changes the arithmetic. Third, the effective FLOP rate tracks the
weight rate, i.e.\ the operator stays arithmetic-bound rather than becoming epilogue bound as
batch grows, and $B{=}256$ ($142.9$\,TFLOPS at $593$\,GB/s) is where that curve meets HBM.

\subsection{Numerical accuracy}

The operator passes 94 of 94 cases and 752 of 752 per-rank checks across default, boundary and
asymmetric routing (including empty and maximally skewed batches); Cake passes 90 of 90 and 720
of 720, skipping the empty-batch cases by convention. Table~\ref{tab:acc} summarizes. The
reference is a dense BF16 MoE computed before quantization; the acceptance threshold is cosine
$0.97$ for the routed variant and $0.99$ for the shared-expert variant, whose residual is also
checked against an independently computed shared branch.

\begin{table}[t]
\centering
\small
\begin{tabular}{@{}llrrrr@{}}
\toprule
Variant & quantity & GLM5.2 & DSV4.1-Flash & threshold \\
\midrule
routed & min cosine vs reference & 0.97610 & 0.97699 & 0.97 \\
 & min cosine vs Cake & 0.99999976 & 0.99999976 & 0.97 \\
 & max $|$GC $-$ Cake$|$ & 0.000000 & 0.000000 & -- \\
 & cases / rank checks & 47/47, 376/376 & 47/47, 376/376 & -- \\
\midrule
fused SE & min cosine vs reference & 0.99999952 & 0.99999958 & 0.99 \\
 & min GC shared residual & 0.99999785 & 0.99999791 & 0.99 \\
 & cases / rank checks & 47/47, 376/376 & 47/47, 376/376 & -- \\
\bottomrule
\end{tabular}
\caption{Accuracy summary. ``min'' is the minimum over ranks and cases.}
\label{tab:acc}
\end{table}

Two facts are worth separating. The $0.976$--$0.977$ cosine of the routed variant against the
BF16 reference is the error of MXFP4 weights and FP8 activations, and Cake reproduces it to the
last printed digit---the two operators agree to $0.99999976$ cosine with a maximum absolute
difference that rounds to zero at this precision. That agreement, not the $0.97$ threshold, is
the meaningful statement: the SM120 scheduling machinery (batched returns, deduplicated reads
and the fused shared expert included) is numerically invisible.

\section{Related Work}

\paragraph{MoE systems.} Tutel~\cite{tutel} introduced adaptive parallelism and a two-dimensional
scheduling policy for MoE; FasterMoE~\cite{fastermoe} and Janus~\cite{janus} attack the
all-to-all by splitting expert computation and communication or by decoupling dispatch from
combine; DeepEP~\cite{deepep} provides highly tuned all-to-all kernels for EP. These systems
optimize the
communication library or the global schedule. GC-MoE~\cite{gcmoe} instead fuses the three phases
of one layer into a single persistent graph and partitions the SMs explicitly, which is the
starting point of this work.

\paragraph{Fused MoE GEMMs.} DeepGEMM's Mega MoE kernel and the fused MoE kernels of that
family---here the FlashInfer~\cite{flashinfer} CuTe DSL W4A8 MegaMoE and Cake MegaMoE---fuse
dispatch, grouped GEMM and combine for FP8 and FP4 on datacenter parts. They are the natural comparison for a
single-layer operator, and they are highly tuned for NVLink-class fabrics and for the
\code{tcgen05} tensor-core path. Our contribution is orthogonal in mechanism and complementary
in spirit: we keep the fusion but re-derive it for a warp-level MMA and a PCIe fabric.

\paragraph{Low-precision microscaling.} The OCP microscaling (MX) formats~\cite{mxspec} define
E2M1/E4M3 payloads with UE8M0 block scales, which is the arithmetic we implement. SM120's
block-scaled \code{mma.sync} exposes them at warp granularity with register scale operands;
CUTLASS~\cite{cutlass} provides the tiling and layout machinery through CuTe DSL, which we use
for layout construction while issuing the mixed-input instruction ourselves.

\paragraph{SM partitioning and overlap.} Green Contexts and programmatic dependent
launch~\cite{pdl} are
the primitives that make intra-layer overlap safe on a single device. Prior overlap work
(communication/compute overlap in data-parallel training, stream-level pipelining) relies on
independent hardware queues; Green Contexts additionally guarantee forward progress between
two co-resident contexts, which is what a spinning consumer needs.

\section{Conclusion}

We presented the SM120 implementation of the GC-MoE operator and the mechanisms that the
architecture forced us to invent: a swap-AB warp-level block-scaled mainloop that keeps the
weight tile invariant while the token extent moves, scale-byte selection through MMA immediates,
two FP4 operand paths with an SMEM-driven choice between them, a run-time adaptive geometry with
SMEM/global/shared handoffs and an explicit register ceiling, streaming expert-granular dispatch
with a batched bulk-pull engine and a 64-bit deduplicating cache, packed routes with mirrored
metadata, full-width BF16 returns that are either coalesced into bulk stores or drained by
resident copiers, and a shared expert folded into the compute role so that fusing it costs no
extra kernel. Together they turn a datacenter design into an operator that is
$1.522\times$ and $1.387\times$ faster than two strong fused baselines on eight
RTX PRO 5000s over PCIe---and $1.388\times$ faster than Cake with the shared expert
fused---at a cross-operator cosine agreement of at least $0.99999976$, i.e.\ a worst-case
cosine distance of $2.4\times10^{-7}$.

\paragraph{Limitations.} The routed advantage narrows to $1.02\times$ at full capacity on
GLM5.2, where the expert working set no longer fits, every operator is HBM bound, and the
copier CTAs are a cost rather than a benefit. That regime is the open problem: it needs either
a return path that does not consume MMA blocks or a weight stream that survives the working
set. The geometry space is hand-derived rather than searched, and the occupancy rewrites of
\S5.1 are tied to a specific ptxas behaviour---they are asserted, but a compiler change forces
revalidation. The design assumes one shared expert and a single-node PCIe fabric; multi-node
fabric behaviour is untested.

\paragraph{Future work.} Three directions look promising: folding the tail of the return path
into the finalize kernel so that no copier blocks are reserved at all, which is what the
$B{=}256$ point above asks for; a learned geometry policy over the same decision chain instead
of the tiered rules of \eqref{eq:tile}; and extending the swap-AB mainloop to a two-stage
accumulation that overlaps GEMM1's epilogue with GEMM2's weight stream at tile granularity.

\clearpage
\appendix
\section{Operator constraints}

The operator validates its geometry when the workspace is constructed;
Table~\ref{tab:limits} lists the accepted range.

\begin{center}
\small
\refstepcounter{table}
\begin{tabular}{@{}ll@{}}
\toprule
Constraint & Supported range \\
\midrule
ranks $R$ & $1 \le R \le 32$ \\
local experts & $E/R$ per rank, $E \bmod R = 0$ \\
top-$k$ & $1 \le k \le \min(32, E)$ \\
hidden $H$ & $512 \le H \le 8192$, multiple of 128 \\
intermediate $I$ & $256 \le I \le 4096$, multiple of 128 \\
activation & SwiGLU (clamp optional) \\
routed recipe & E4M3 activations $\times$ MXFP4 weights, block $(1,1,32)$ \\
shared weight recipe & MXFP8, $K$-group 32 or 128, packed int32 UE8M0 scales \\
routed weight payload & int8 $[E, K/128, N, 64]$, scales int32 $[E, N, K/128]$ \\
shared weight payload & int8 $[K/128, N, 128]$, scales int32 $[N, K/(4\,\text{gran})]$ \\
compute partition & $\ge 32$ SMs, even, $\ge \ell + 1$ co-resident CTAs \\
\bottomrule
\end{tabular}
\par\smallskip
\parbox{\linewidth}{\small Table~\thetable: Supported operator geometry. Payload shapes are
the packed MXFP4 tile form produced by the weight transform; scales are MN-major.}
\label{tab:limits}
\end{center}

\paragraph{Reproducibility notes.} The dispatch/compute split is controlled by
\code{GC\_MOE\_SM120\_\allowbreak DISPATCH\_SMS} (default 6), balanced workqueues by
\code{GC\_MOE\_BALANCED\_WORKQUEUES}, and primary-context roles by \code{GC\_MOE\_PRIMARY\_CTX}.
The geometry of Table~\ref{tab:geom} is fully determined by $(R, E, k, B, H, I)$ and the device
SMEM capacity, so the configuration used at each of the 18 points can be regenerated without
timing instrumentation.

\end{document}